\documentclass[trackchanges]{aastex701}
\usepackage{amsmath}

\usepackage{enumitem,booktabs}

\newcommand{\hetg}{{\small {\it Chandra}/HETG}}

\newcommand{\NH}{$N_{\rm H}$}

\newcommand{\NHtot}{$N_{\rm H}^{\rm tot}$}

\newcommand{\grs}{{\small GRS\,1915+105}}
\newcommand{\gx}{{\small GX\,13+1}}
\newcommand{\fu}{{\small 4U\,1630-472}}

\begin{document}

\title{Stratified Wind Geometries of GRS 1915+105: Density Profiles and Radii}

\author[0000-0003-2535-6436]{Noa Keshet}
\affiliation{Department of Physics, Technion
Haifa 32000, Israel}
\email[show]{noa.keshet@campus.technion.ac.il}  

\author[0000-0001-9735-4873]{Ehud Behar} 
\affiliation{Department of Physics, Technion
Haifa 32000, Israel}
\email{}

\author[0000-0003-2869-7682]{Jon M. Miller}
\affiliation{Department of Astronomy, University of Michigan, Ann Arbor, MI 48109, USA}
\email{}

\author[0000-0002-8247-786X]{Joey Neilsen}
\affiliation{Villanova University, Department of Physics, Villanova, PA 19085, USA}
\email{}

\begin{abstract}
The low-mass X-ray binary \grs\, exhibits dramatic spectral and flux variability, often accompanied by highly ionized outflows. 
In this work, we analyze four archival \hetg\, observations spanning over a decade, which vary greatly in their ionizing spectrum from soft (low inner disk temperature) to hard (high inner disk temperature). We measure the wind properties, using the insightful tool of the absorption measure distribution (AMD, a distribution of column density with ionization). While the soft epoch reveals a steep, monotonically rising AMD, hard epochs feature a distinct turnover at high ionization parameter ($\log\xi>3.3$). 
This is the first time we observe a negative AMD slope in an outflow, which we interpret as a localized density inversion, which could point to radiation-pressure compression zones. 
We utilize the AMD shape and total column density measurements to calculate the density profile and distance of the wind from the central source. 
For the hard states, we obtain densities of $10^{12}-10^{13}$ cm$^{-3}$ at radii of $10^{10}-10^{11}$\,cm. Such measurements are usually not directly possible in X-ray spectroscopy. Here we constrain the inner radius and density of the outflow to a factor of two.
The high-$\xi$ values up to $\log\xi\sim6.5$ with no turnover in the soft state imply a smaller radius of a few $10^{8}$\,cm, which is in tension with the slow velocities observed.
We demonstrate that this is a generic tension between compact launch radii and sub-escape velocities of high-column, high-ionization outflows observed in stellar-mass black hole binaries.

\end{abstract}



\section{Introduction}

\grs\, is a unique low-mass black hole X-ray binary that has shown intense variability over  almost three decades of X-ray observations. \grs\ is a wide binary with a semi-major axis of $\simeq7\times10^{12}$cm \citep{Miller2025}, a $10.1\pm0.6\,M_{\odot}$ black hole with a $0.5\pm0.3\,M_{\odot}$ companion \citep{Steeghs2013}. It has been observed in several outbursts featuring blue-shifted absorption lines, indicating an outflow. These outbursts usually feature bright continuum emission, modeled as a multi-temperature black-body from the accretion disk, which is in turn the main ionizing source. There is strong evidence for a magnetic launching mechanism, based mainly on the launching radius of the wind at close proximity to the black hole \citep{Miller2015, Miller2016, Ratheesh2021}. \citet{Miller2016} find wind launching radii of $r\simeq10^{2-4}$ GM/c$^{2}$ and density values of $n\simeq10^{13-16}$\,cm$^{-3}$, for a soft, high state. This is in agreement with the results of \citet{Zoghbi2016}, that constrain the wind launch radius to $290-1300\,r_g$, based on the changes in wind velocity and absorbed flux as observed with \textit{NuSTAR} and \textit{Chandra} in a joint observation. \citet{Ueda2009} placed the wind further from the black hole, at $r\sim(2-6)\times10^{11}\,$cm, or $\sim(1-3)\times10^5\,r_g$, assuming a density of $n=10^{12}$cm$^{-3}$.
Similarly, there are estimations of the density and launch radius for the hard, low state. \citet{Miller2020} model the outflow with two zones, and constrain the upper limit of the launching radius of each one. The "inner" absorption zone is highly ionized and constrained to $r\leq9.7\times10^9$cm, or $r\leq5.2\times10^3\,GM/c^2$. The "outer" absorption zone is constrained to $r\leq1.6\times10^{11}$\,cm, or $r\leq8.8\times10^4\, GM/c^2$.  \citet{Neilsen2020} place the wind at a ~few$10^{11}$cm from the black hole, with a density of $\sim10^{12-13}$\,cm$^{-3}$. These values are estimated based on an analytical description of the absorber and allow for several different scenarios regarding its density profile and geometry. 
All of the works have only estimations or ranges for the density and radii measurements, and in many of the cases there are additional specific assumptions made. The most recent X-ray high-resolution observation of \grs\ was taken with the \textit{XRISM} telescope \citep{Miller2025}. It was observed in a highly obscured state, with only emission lines in the spectrum. They estimate the origin of a broad Fe XXV line as close as $r\simeq3\times10^3\,GM/c^2$, compatible with the location of a neutral Fe $K_\alpha$ line. Taking the Roche lobe radius and assuming the accretion disk fills about two thirds of it, the outer disk extends to $r_{\rm out} = 1.8\times10^6\,GM/c^2$, which is large enough to produce the narrow emission lines observed. \citet{Miller2025} suggest a picture where this state of \grs\ is a result of obscuration by the irradiated outer disk, due to a warped, precessing disk that brought the outer disk into the line of sight.


To characterize the appearance of many absorption lines in the spectrum one defines the ionization parameter: 
\begin{equation}
    \label{eq_xi}
    \xi = \frac{L}{nr^2}
\end{equation}
\noindent where $L$ is the ionizing luminosity, $n$ the H number density, and $r$ the distance from the ionizing source. $\xi$ represents the balance between photo-ionizing flux ($\propto L/r^2 $) and recombination ($\propto n$) rates. A useful tool that provides insight into the physics of the wind is the absorption measure distribution, the AMD. 
The AMD is the distribution of the hydrogen column density $N_{\rm H}$ as a function of $\xi$ \citep{Holczer2007}:

\begin{equation}
    \label{eq_AMD}
    AMD = \frac{dN_{\rm H}}{d\log \xi}
\end{equation}

In previous work \citep{Keshet2025} we utilized the AMD as part of the process of measuring the elemental abundances in the system, thus gaining insight to the black-hole progenitor. 
The AMD was found to follow a similar power-law for the three X-ray binaries \grs , \fu , \gx . The power-law slope was steep, indicating high column density at high ionization states. As part of the sample, only one outburst of \grs\, was included, and it shared similarities with outflows of the other binaries. In this work, we aim to explore the changes in AMD as the spectrum of \grs\, evolves over several flux levels. This serves to highlight how the change in continuum affects the outflow. We further take advantage of the AMD analysis to calculate the location, density profile, and size of the wind. Integrating over Eq.\,\ref{eq_AMD} results in the total column density in the wind. The shape of the AMD itself is connected to the density profile $n(r)$ through the ionization parameter definition. We expect the AMD method to constrain the outflow properties since it provides a complete description of the ionization structure. Thus, we are able to calculate values for the density and radii of the wind, which so far have only been estimated to within orders of magnitude or given limits by using physical arguments.

\section{Observations and Data}

\begin{figure}
    \centering
    \includegraphics[angle=270, width=0.9\linewidth]{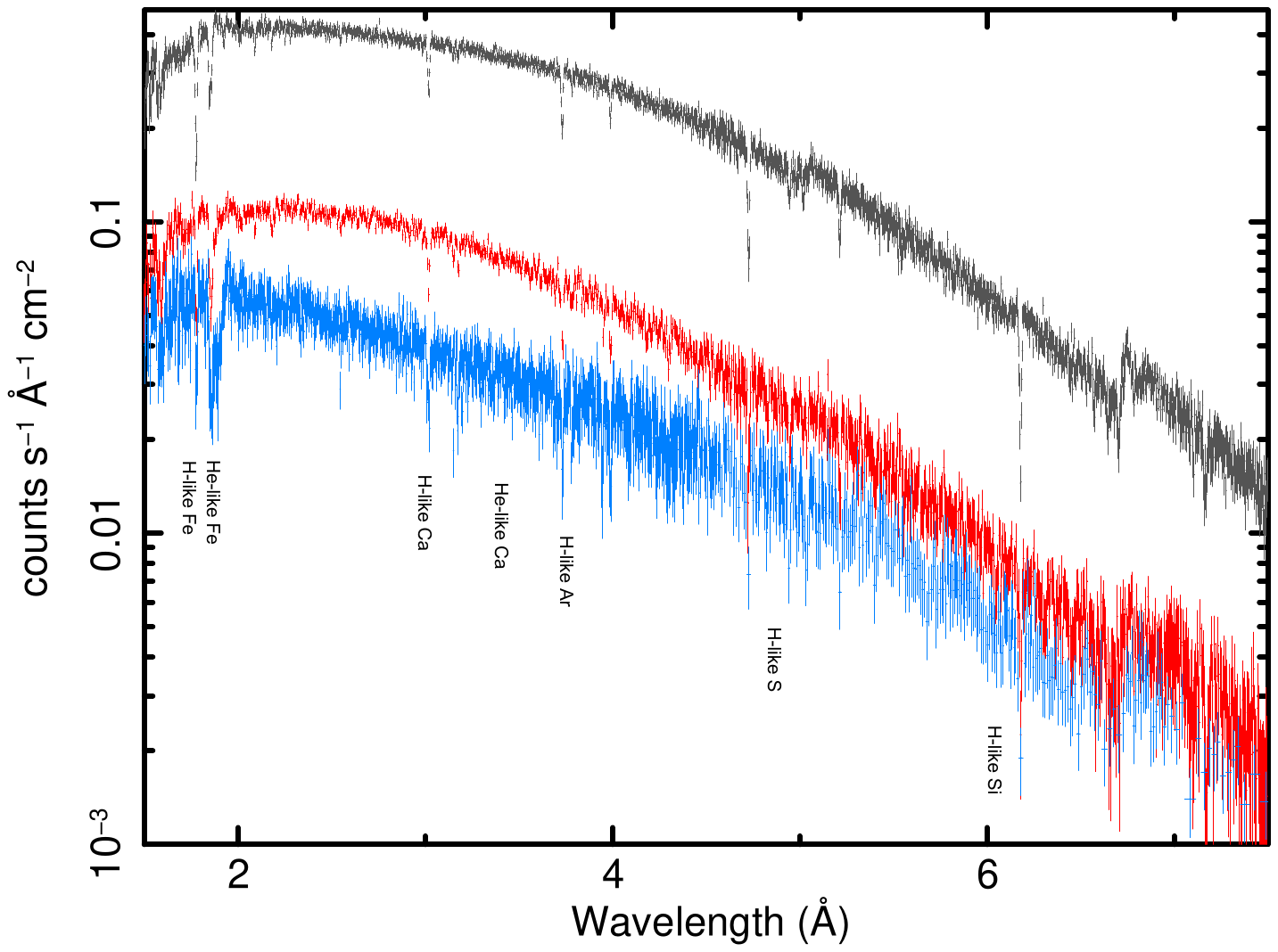}
    \caption{\textit{Chandra}/HETG observations of the four observations used in the paper. The figure presents the $1.5-7.5 $\AA\, part of the HEG first order. In gray - observation 7485, in red - combined observations 23435 and 24663, in blue - observation 22213.  }
    \label{fig:3spectra}
\end{figure}

\begin{figure}
    \centering
    \includegraphics[angle=270, width=0.3\linewidth]{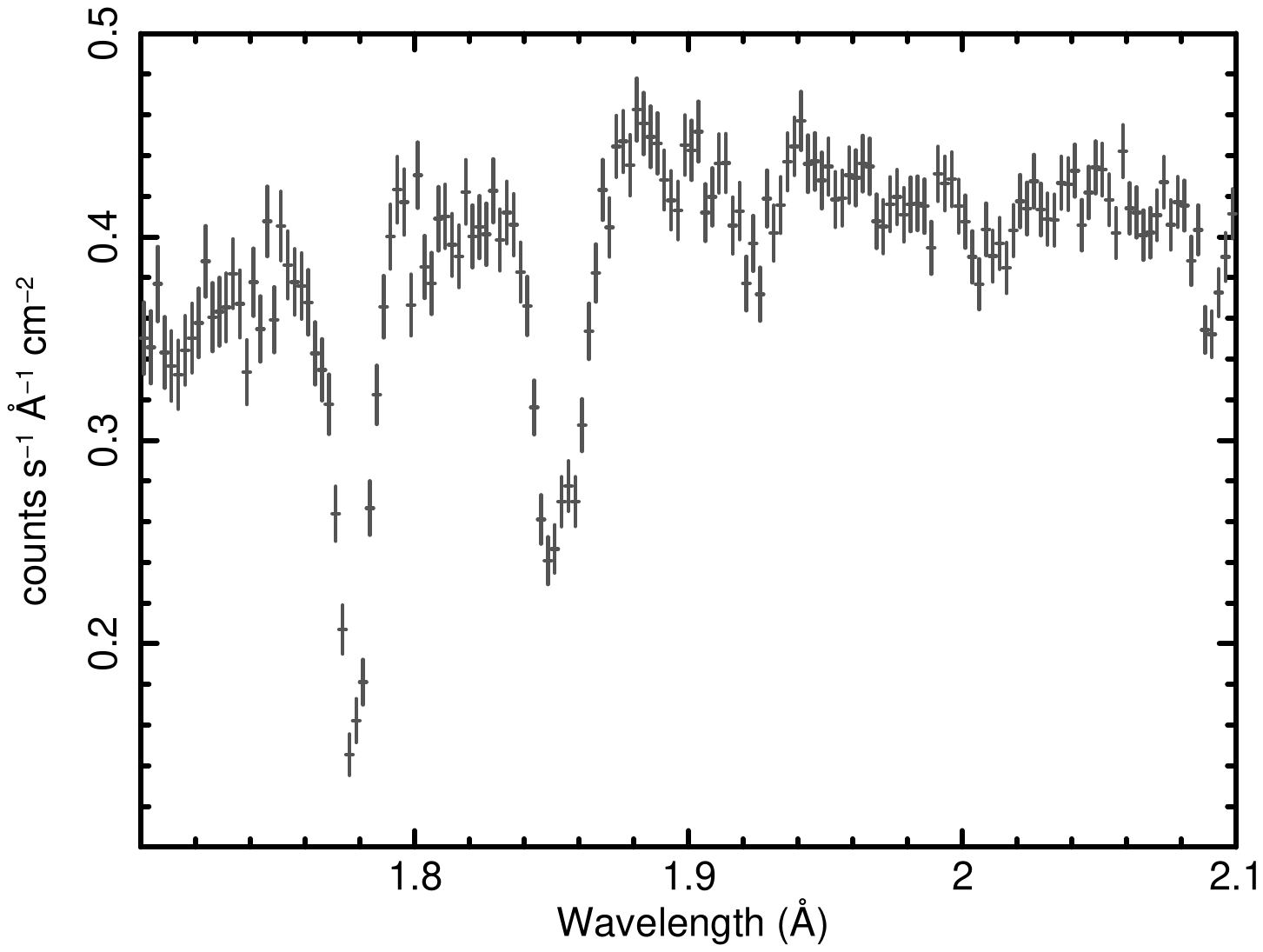}
    \includegraphics[angle=270, width=0.3\linewidth]{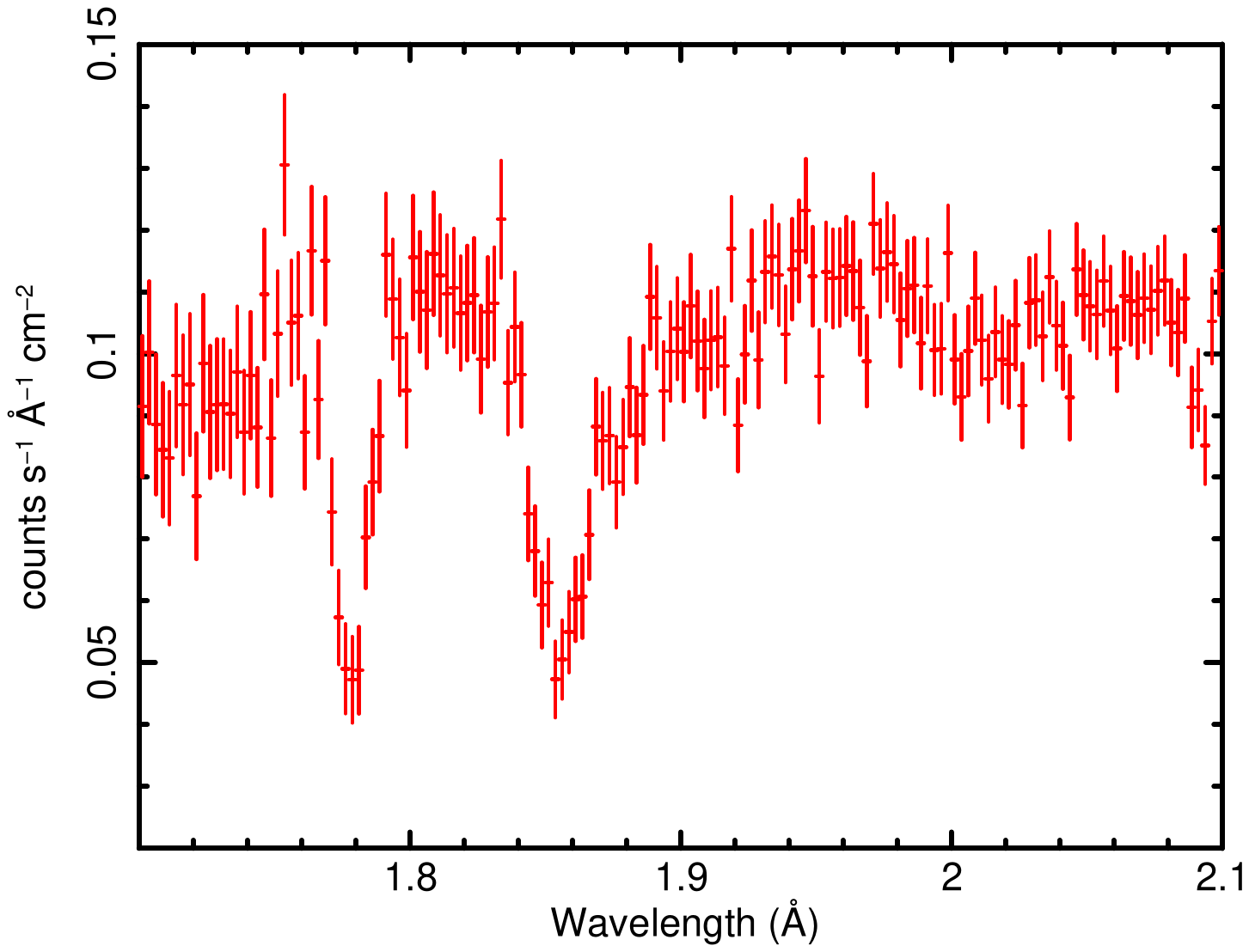}
    \includegraphics[angle=270, width=0.3\linewidth]{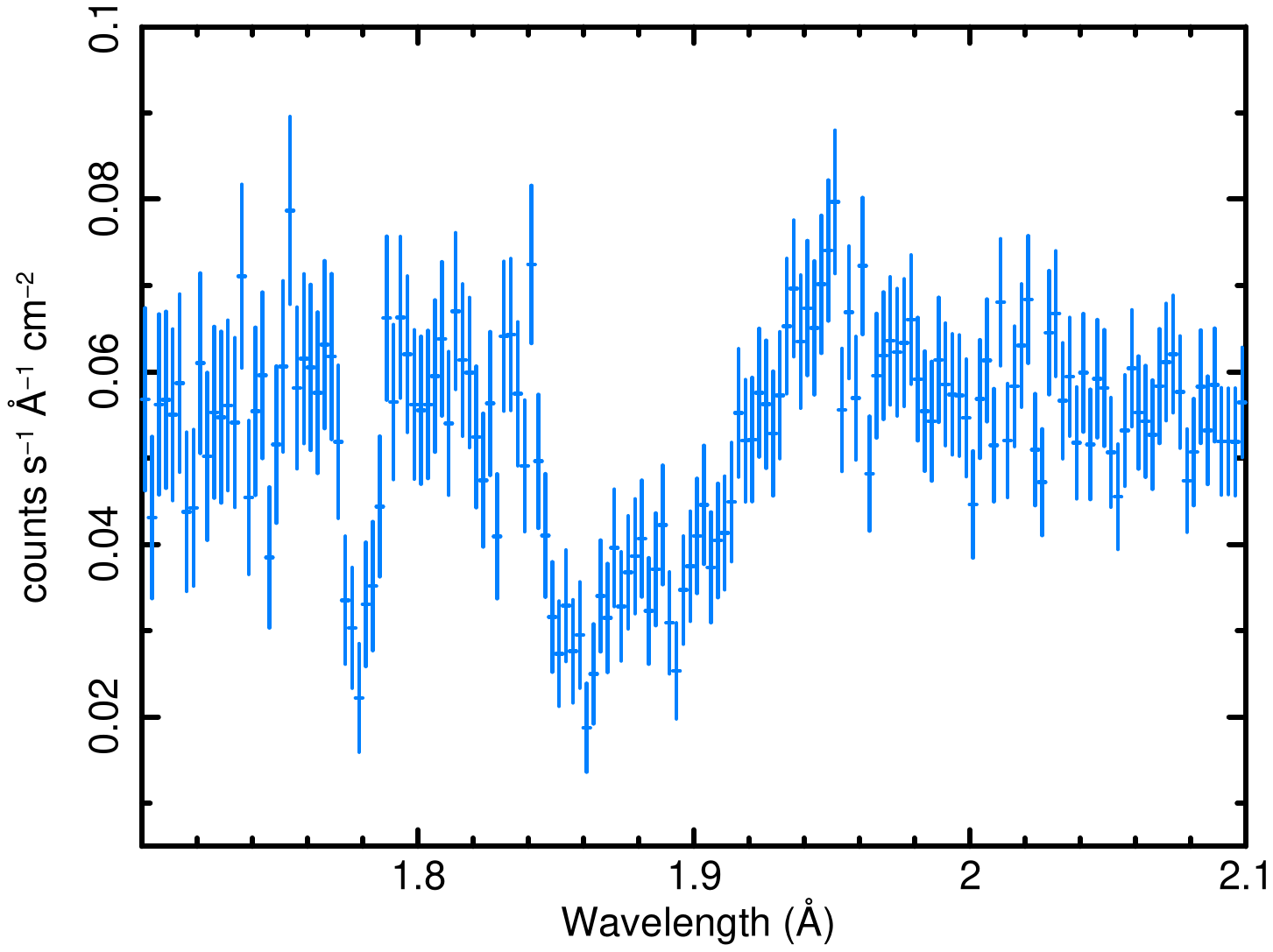}
    \caption{Zoom into the Fe K region of observations 7485, 23435 and 24663 combined, 22213 in this order from left to right. the transition from a dominant H-like (1.78) trough to dominant He-like (1.85) and lower charge states above 1.9\AA\, can be seen.}
    \label{fig:Fe}
\end{figure}

In this work we examine four archival observations of \grs\, taken with the \hetg\, over more than a decade. All archival observations were downloaded from TGCat. Other \hetg\ observations of \grs\ did not have enough useful absorption lines, except one (obsid 16711) that did have absorption lines, but was acquired in continuous clocking mode. The details for the four observations we deem useful appear in Table\,\ref{tab:obs}. All of these observations present clear outflows with absorption lines of at least seven different elements. An additional important criterion for the AMD analysis is the presence of lines from pairs of H-like and He-like ions of the same element. In all observations at least five such pairs were clearly identified. 
Figure\,\ref{fig:3spectra} presents the four first order HEG grating spectra, plotted in log scale the $1.5-7.5$\,\AA\, range. The two observations from 2021 were taken on consecutive days and present similar spectra, therefore they are plotted grouped together and analyzed as one spectrum. Figure\,\ref{fig:Fe} focuses on the changes of the Fe K region across observations. 

\begin{deluxetable}{lccccc}
\tabletypesize{\scriptsize}
\tablewidth{0pt}
\tablecaption{\hetg\ Observations of \grs\, Used in the Present Work
}
\label{tab:obs}
\tablehead{
\colhead{Observations ID} 
& \colhead{Start Date} 
& \colhead{Exposure} 
& \colhead{Previous HETG} \\
\colhead{}
& \colhead{}
& \colhead{(s)}
& \colhead{Absorber Analysis}
}
\startdata
7485  & 2007 Aug 14 & 48759    & \cite{Ueda2009,Miller2015} \\
      &             &          & \cite{Ratheesh2021,Keshet2025} \\
22213 & 2019 Apr 30 & 29085    & \cite{Miller2020}\\
23435 & 2021 Jul 14 & 24503    & \cite{Parra2024}\\
24663 & 2021 Jul 15 & 23504    & \cite{Parra2024}\\
\enddata
\end{deluxetable}

\section{Method}

Measuring elemental abundances in all observations provides a self-consistency check of our method, presented in detail in \citet{Keshet2024, Keshet2025}. Since for an accurate abundances measurement one needs to reconstruct the ionization distribution of the outflow, this creates a chance to compare this important property in different outbursts of the source. 

\subsection{Ionic Column Densities and Reconstruction of the AMD}

The equivalent width $EW$ of an absorption line is the integral over its profile $EW = \int (1-e^{-\tau(E)})dE$, where $\tau (E)$ is the optical depth. The $EW$ can be measured directly from the spectrum. Its curve of growth $EW(N_{\rm ion})$ yields $N_{\rm ion}$.
The measured absorption $EW$s thus represent the ionic column densities, which can be expressed as:
\begin{equation}
    \label{eq_Nion}
    N_{\rm ion} = A_{\rm Z} \int f_{\rm ion}(\xi) \frac{dN_{\rm H}}{d\log \xi} d\log \xi
\end{equation}

\noindent where $N_{\rm H}$ is the hydrogen column density, $A_{\rm Z}$ is the elemental abundance with respect to H, $\xi$ is the ionization parameter (Eq.\,\ref{eq_xi}), and $f_{\rm ion}(\xi)$ is the ionic fractional abundance.
The $f_{\rm ion}(\xi)$ are calculated using the Cloudy code \citep{Ferland2013} and are individually computed for each observation, based on the ionizing continuum. The best fitted continuum parameters for three spectra are listed in Table\,\ref{tab:continuum}. 
Ionizing continua are comprised of a multi-temperature disk black-body and a hard X-ray power-law. The disk black-body is characterized by the inner disk temperature, $T_{in}$.
The maximal disk temperature (hardness) varies dramatically between observations, which considerably affects the $\xi$ at which the ions form. The non-thermal power law component also varies between observations, but is subdominant. Specifically, the flat slope in 23435+24663 has a negligible normalization. 
The ionic column densities are measured with the ion-by-ion fitting code \citep{Peretz2018}, which implements the curve of growth method for all lines and simultaneously fits for the $N_{\rm ion}$ values and a global velocity width. The ionic column densities were also measured using the Ionabs code in Xspec \citep{Tomaru2020} to validate consistency. 

\begin{deluxetable}{lccccccccc}
\tabletypesize{\scriptsize}
\tablewidth{0pt}
\tablecaption{Best-fit continuum parameters for each observation 
}
\label{tab:continuum}
\tablehead{
\colhead{}  
& \colhead{Galactic Absorption}
& \multicolumn{2}{c}{Disk black body}
& \multicolumn{2}{c}{Power-law} \\
\colhead{Observation} 
& \colhead{$N_{\rm H} (\times10^{22}/\rm cm ^{2})$}
& \colhead{$kT $ (keV)}
& \colhead{Norm}
& \colhead{Photon index}
& \colhead{Norm $\rm\,photons/keV/ cm^{2}$}
}
\startdata
{7485}& $5.24\pm0.07$ & $1.35\pm0.04$ & $133\pm23$ & $2.03\pm0.08$ & $4.4\pm0.07$ \\
{23435+24663} & $5.4\pm0.1$ & $1.86\pm0.02$ & $19.1\pm0.1$ & $1.1\pm0.1$ & $0.022\pm0.005$ \\
{22213} & $5.0$ fixed & $2.52\pm0.02$ & $2.7\pm0.2$ & $2.7\pm0.4$ & $0.24\pm0.05$ \\
\enddata

\end{deluxetable}

\begin{figure}
    \centering
    \includegraphics[width=0.9\linewidth]{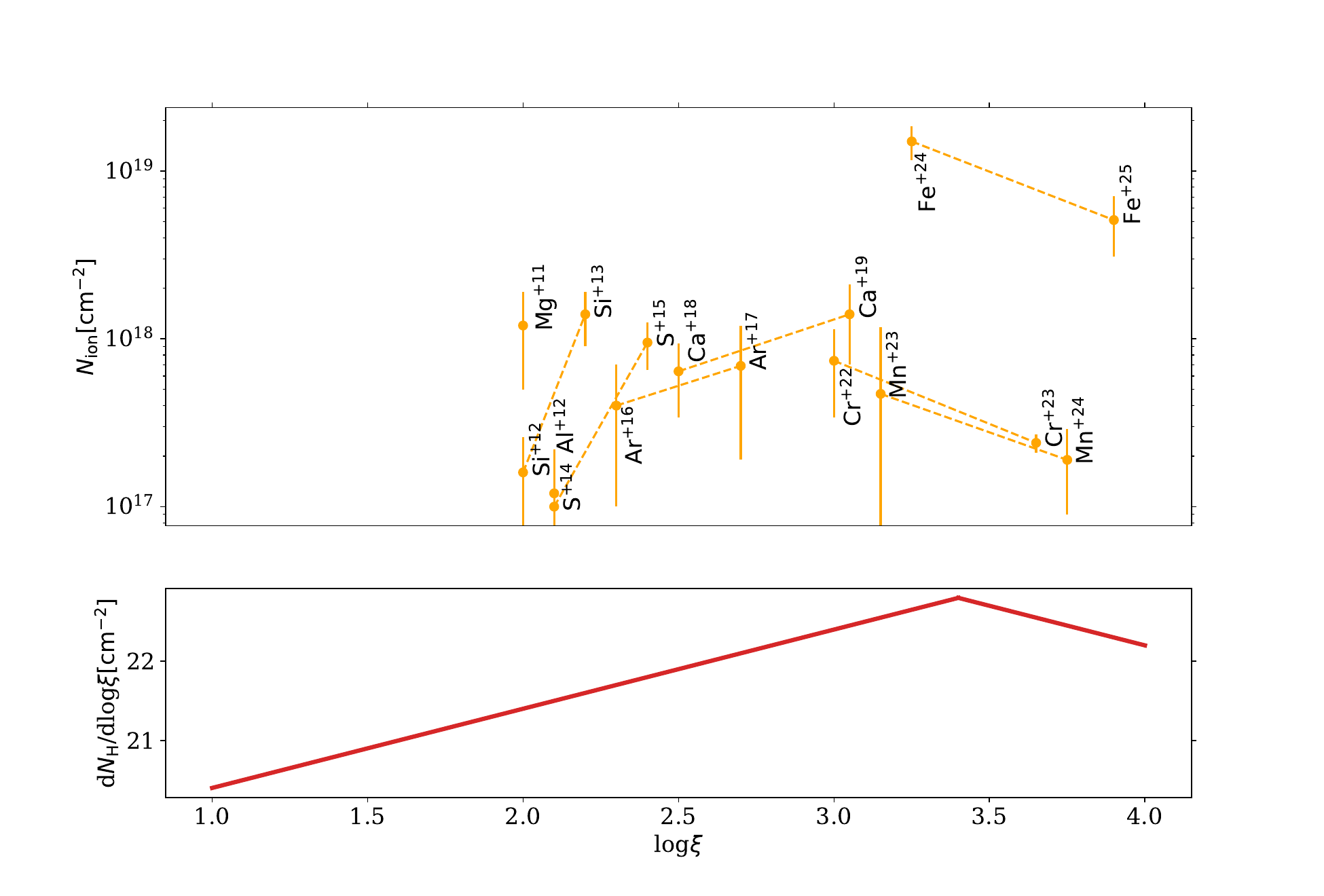}
    \caption{(\textit{top panel}) Distribution of measured ionic column densities $N_{\rm ion}$ in the \grs\ outflow of observations 23435+24663, each plotted at its $\xi$ of maximal formation. 
    This outlines the overall distribution of $N_{\rm H}$ with $\xi$, showing a gradual increase, and then a turnover. The dashed lines represent the local slopes from H-like and He-like ion pairs, which are used to reconstruct the continuous AMD (see text), plotted in the (\textit{bottom panel}).  
   Since there is no H in the spectrum, $f_{ion} (\xi)$ and $A_Z$ of H-like Fe are used to scale the absolute AMD values in terms of \NH (Eq.\,\ref{eq_Nion}).}
    \label{fig:2324AMD}
\end{figure}

The shape of the AMD is achieved based on the following assumptions: 1. Different ions of the same element must result in the same $A_{\rm Z}$ measurement. Therefore a difference in ionic column densities of the same element represent a general behavior of \NH($\xi$). 2. The AMD is a power-law or consists of several power-law segments tailored together. This is assumed for simplicity. 3. Breaking the AMD into segments is only done when necessary and the single power-law AMD does not provide the above requirements, attempting to reach the simplest overall solution that still produces consistent abundances.  The process of the reconstruction of the AMD is demonstrated for the joint analysis of observations 23435 and 24663 in Figure\,\ref{fig:2324AMD}. In the top panel all of the measured ionic column densities are plotted at their peak-$\xi$ formation value. However, the calculation of ionic column densities (Eq.\,\ref{eq_Nion}) includes their fraction at both ends down to 10\%, hence the broader range of the bottom panel. Our estimate of the AMD attempts to best match the measured $N_{\rm ion}$ values in Eq.\ref{eq_Nion}. The resulting AMD is shown in the bottom panel scaled to the H-like Fe ionic column density assuming the Fe abundance is solar.


\subsection{Density and Radii Calculations}
\label{sec:density calculation}

We utilize the shape of the AMD to find a consistent solution for the density profiles of the outflow. Integrating over the entire AMD gives the value of \NHtot\, i.e. the integral over Eq.\,\ref{eq_AMD}. \NHtot can also be calculated integrating over the profile density - $N_{\rm H} = \int ndr$. Following \citet{Behar2009}, we translate the AMD slope to a density profile, $n(r)$. Each segment of the AMD corresponds to a segment of the wind between two radii, with a density profile of the shape $n\sim r^{-\alpha}$. Combining the resulting integrals (two or three, depending on the observation) and integrating over the entire length scale of the outflow, sums up to the measured \NHtot. We reach a set of equations created by this requirement \ref{eq_n0}, where each separating radius ($r_i$) is associated with the relevant break ($\xi_i$) found in the AMD (Eq.\,\ref{eq_r}). 


\begin{equation}
    N_{\rm H} = n_0 \left ( \int_{r_0}^{r_1}(\frac{r}{r_0})^{-\alpha _1} dr+ \int_{r_1}^{r_2}(\frac{r}{r_0})^{-\alpha _2 } dr + .. \right)
    \label{eq_n0}
\end{equation}

\begin{equation}
    r_i = \sqrt{\frac{L}{n_0 \xi _i}} 
    \label{eq_r}
\end{equation}


\noindent Note that the relation between $\alpha$, the density profile slope, and $a$, the AMD slope ($AMD\sim\xi^a$) can be different based on the assumption of the size of the outflow \citep{Behar2009}. In the density and radii calculations we use a constant value for the luminosity. This is a simplifying assumption that may be too crude. Since these outflows carry large column densities, it can lead to significant absorption of the radiating source luminosity, which requires a more delicate approach in the calculation. This is beyond the scope of this work.

Due to the a-priori unknown shape of the AMD and the reconstruction method, there are no standard deviation uncertainties calculated for the slopes of the different segments. To get an estimation of the uncertainties of the density and radii results, we first test different limits for the slopes of the AMD segments, demanding the elemental abundances agree within the formal $N_{\rm ion}$ uncertainties between ions of the same element (Eq.\,\ref{eq_Nion}). This is done by taking each segment (one, two or three) of the AMD and varying its slope until reaching a limit where the resulting AMD does not give the same abundance value for the ions of the same element. We thus obtain the shallowest and steepest slopes allowed. There are no a-priori limits on the range of $a$ or \NHtot .
The uncertainty of the $\log\xi$ value of the breaking point was not explored. Subsequently, we solve the set of equations (Eqs.\,\ref{eq_n0},\ref{eq_r}) with the above extreme values of $a$ and \NHtot. We obtain a range of $n_0$ values and radii, which represent the uncertainty of these values. The results of this elaborate process are detailed in the next section.

\section{Results}
\label{sec:Results}

\subsection{AMD Shape and \NHtot}
\label{sec:AMD shape}

\begin{figure}
    \centering
    \includegraphics[width=\linewidth]{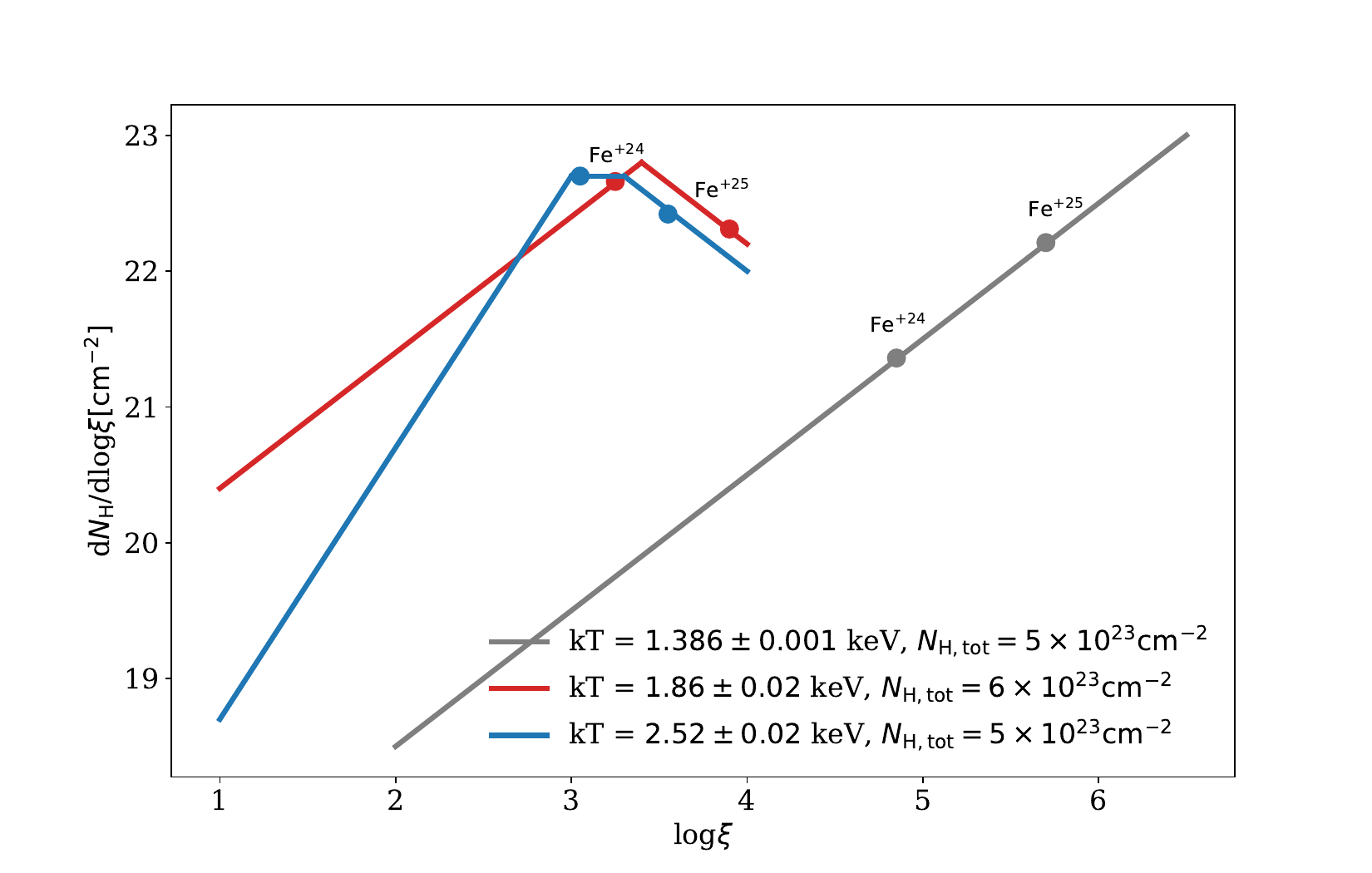}
    \caption{The AMD reconstructed from all three epochs of GRS. Each color correspond to a different observations: gray for obsId 7485, red for 23435+24663, blue for 22213. For each observations the temperature of the disk black body is mentioned, as well as the total column density up to the decided cutoff, see Sec\,\ref{sec:AMD shape} for details. On each AMD the H-like and He-like Fe points are marked, at their maximal $\xi$ formation.  }
    \label{fig:AMDs}
\end{figure}

The three different AMDs reconstructed for each of the spectra are plotted in Figure\,\ref{fig:AMDs}. The colors correspond to the colors of Figure\,\ref{fig:3spectra}: gray for observation 7485, red for the joint analysis of 23435 and 24663, and blue for 22213. The high cutoff for each of the AMDs is important for the measurement of \NHtot, calculated by integrating over the distribution. This cutoff is determined based on a uniform criterion, where the fractional abundance of H-like Fe comes down to $20\%$. 
The location of maximal formation for H-like and He-like Fe are plotted in Figure\,\ref{fig:AMDs} for each AMD. This demonstrates how the change in SED, mainly the temperature of the disk black-body, affects the ionization distribution. As the disk temperature rises, the photon population comprising $L$ shifts towards higher energies, therefore ionization species form at lower $\xi$ values, see Eq.\,\ref{eq_xi}. The non-linearity of the shift in $\xi$  reflects the non-linearity of the number of ionizing photons in a disk black body SED. We stress that the abundances obtained in the AMD reconstruction are fully consistent with previous measurements of abundances in \grs\ \citep{Keshet2025, Miller2025}.

The AMD shows dramatic changes between observations. The shape of a single, steep power-law was found to be common in X-ray binaries outflows \citep{Keshet2025}. For the first time in such outflows the AMD is found to have a break, with negative slopes at high $\xi$. A negative AMD slope leads to an inversion in the density profile, see further details in the discussion. The origin for this result is clear when examining the high-Z element pairs in the top panel of Figure\,\ref{fig:2324AMD}. Observing weaker H-like lines in comparison with the He-like lines (see Figure\,\ref{fig:Fe}) generally means there is less column density at higher ionization, which translates to a negative slope in the AMD. This behavior is robust and is not dependent on the specific photo-ionization fractional abundance assumed. Since observations 23435, 24663 and 22213 are in a lower flux state, it is possible that we observe only the absorbed ionizing continuum and not what the outflow is actually exposed to. Therefore, we repeated the process of the AMD reconstruction using the fractional abundances calculated for the highest flux state from observation 7485. We obtain the same AMD shape, but horizontally shifted along the $\xi$-axis. This is due to the mentioned relation to the disk black-body temperature. 

Integrating the AMD up to the $\log\xi$ cutoff values yields a \NHtot for each observation of $5\times10^{23}\,\rm cm^{-2}$ for 7485, $6\times10^{23}\,\rm cm^{-2}$ for 23435+24663 , and $5\times10^{23}\,\rm cm^{-2}$ for 22213. For observation 22213 this \NHtot\, measurement is comparable to previously measured results \citep{Miller2020}. Since the AMD is steeply rising in observation 7485, it is possible there is a significant amount of entirely ionized material which will contribute to the total column, making the source Compton-thick. Such a state could have implications on the continuum observed, where the wind is obscuring the inner disk \citep[see][]{Neilsen2016}.

\subsection{Density and Radii}

\begin{figure}
    \centering
    \includegraphics[width=0.45\linewidth]{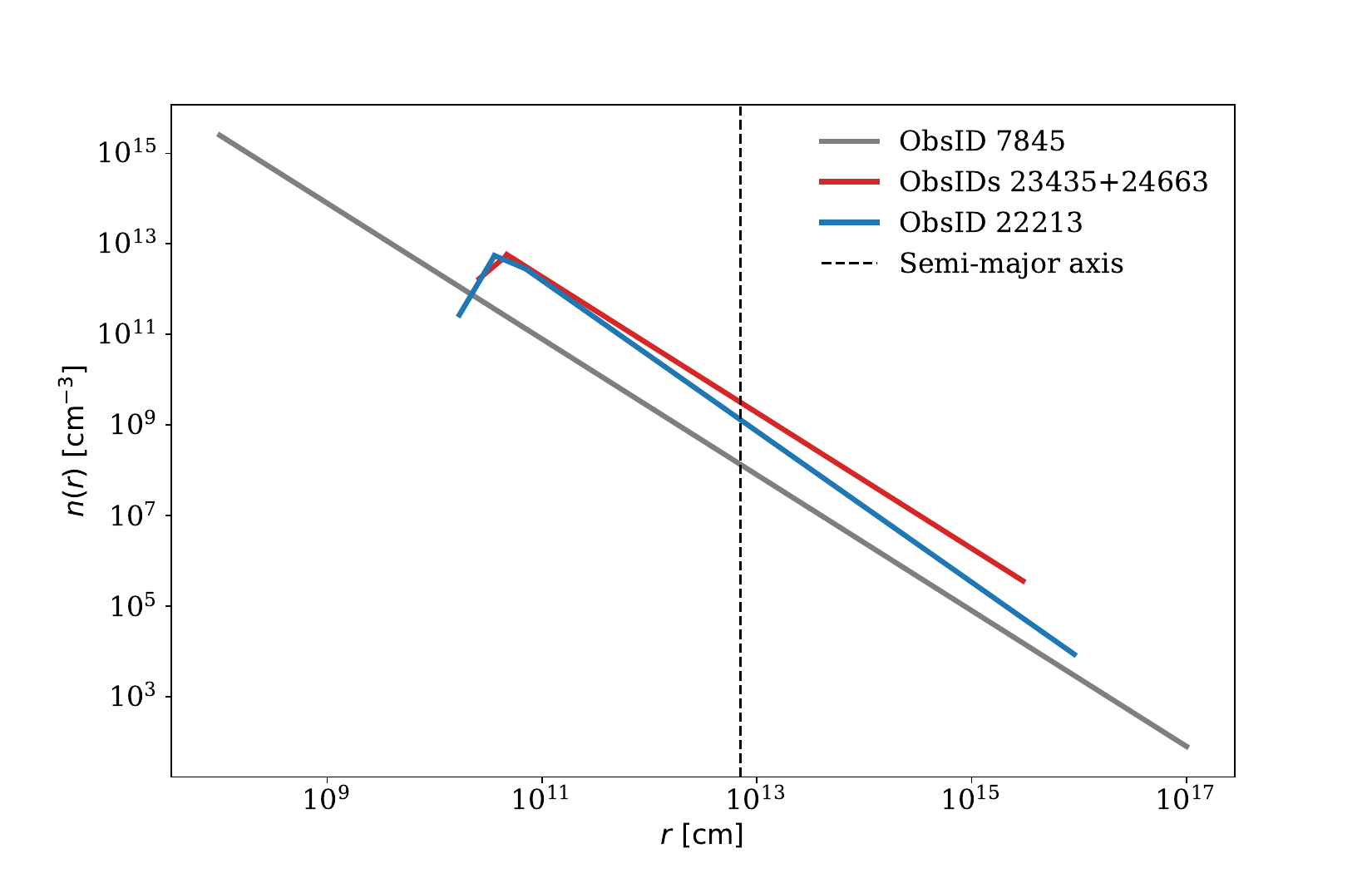}
    \includegraphics[width=0.45\linewidth]{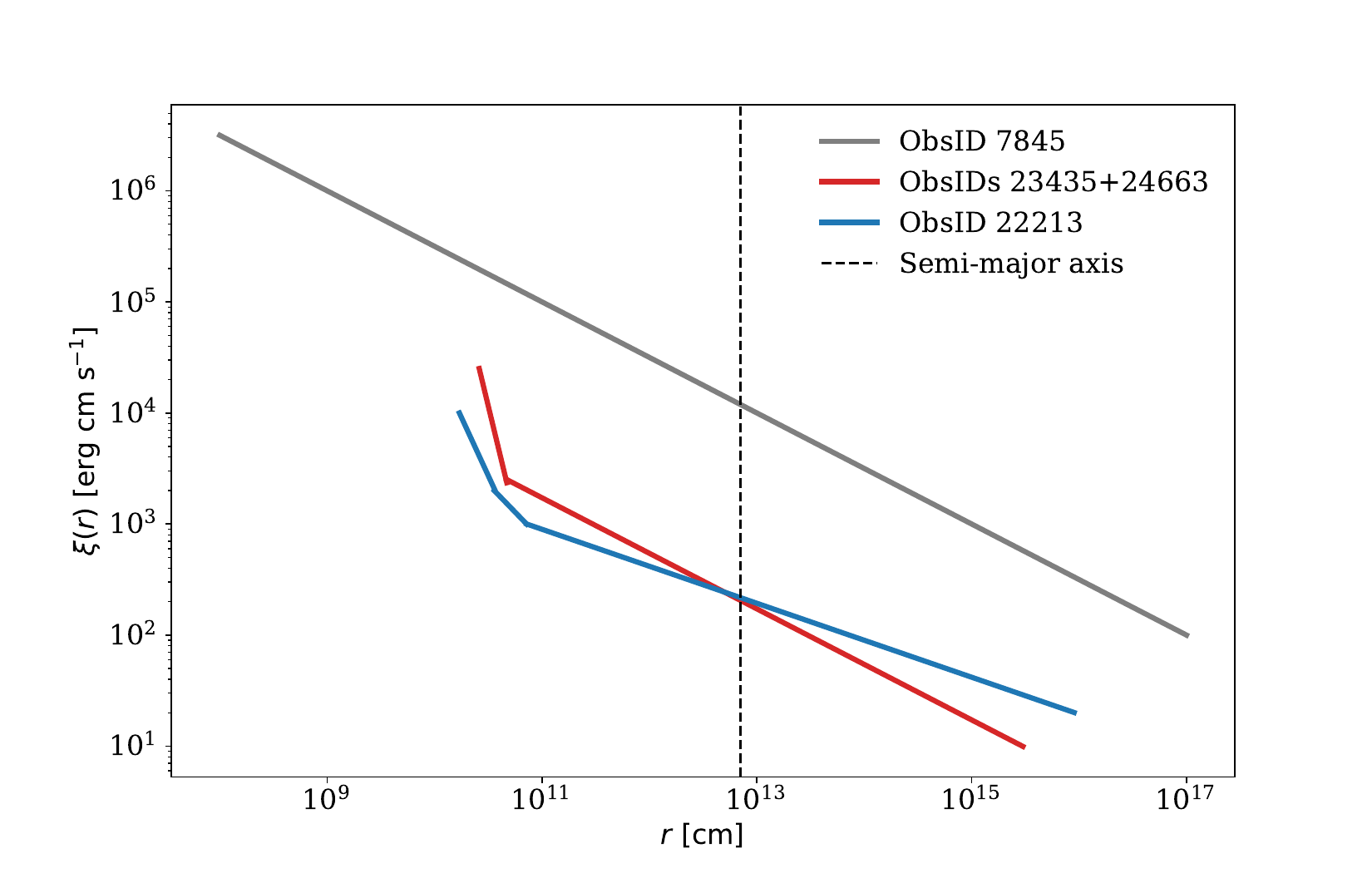}
    \caption{\textit{Left} The calculated density profile as a function of the radius, for each of the observations. \textit{Right} The calculated ionization paramtere as a function of the radius for all observations. }
    \label{fig:density}
\end{figure}

The density and radii calculation results are listed in Table\,\ref{tab:density}. The table details the range for each measurement. How this range was calculated is explained in Sec\,\ref{sec:density calculation}. For each quantity, the $\xi$ at which it is defined is mentioned in the footnotes below the table. Since all density and radii measurements are translated from the reconstructed AMD, the radii $r_i$ are defined according to the start, end and break points of the AMD. Notice that for observation 22213 there is a plateau at the mid-ionization section, therefore there are two radii in the $r_{\rm peak}$ column that correspond to the edges of the plateau. The density profile and the behavior of the ionization parameter as a function of the radius are plotted in Figure\,\ref{fig:density}.

For the translation between the measured $a$ slope of each AMD segment to the $\alpha$ slope of the density profile, one needs to choose the length scale of the outflow. The outflow could be on large scales, $r\sim \Delta r$, or on small scales  $\delta r$, where the density varies over a short range \citep{Behar2009}. As seen in Figure\,\ref{fig:2324AMD}, the range of $\xi$ in which the AMD rises is by far wider than the declining region. Consequently, for the rising segments of the AMD, as well as the plateau, we referred to the large scale solution, while for the dropping segment we prefer the small scale behavior. This provided the most self-consistent results. We elaborate on the meaning of a negative AMD slope and a small scale wind in the Discussion. The values for $r_{\rm max}$ are several orders of magnitudes larger than the semi-major axis of the system. These value are directly connected to the lower $\xi$ cut-off of the AMD. Since the lower-Z elements, that peak at these lower $\xi$ values, are much weaker than the more ionized species, the exact point for this cut-off is somewhat flexible. This part of the AMD contribute a small fraction of the total column, therefore we are not sensitive to an exact choice of $\xi$. Choosing a higher $\xi$ would result in smaller values of $r_{\rm max}$.   

The density and radii calculations for observation 7485 result in extremely low radii, which are hard to constrain within a complete physical picture. For a black hole of this size we would not expect to find such slow winds at such close proximity of $~10^8$cm. This is a result of the steep AMD with no break up to a very high $\xi$ point of 6.5. We tested a slightly more complex AMD to better understand its behavior and connection to the density profile. An alternative AMD that still give a consistent solution for the measurements of observation 7485 is a power-law with a slope of $a=1$ up to $\log\xi=5.7$, with a plateau at high $\xi$ in the range $\log\xi=5.7-6.5$ . This broken AMD provides a more physical solution, where the density and radius at $\log\xi=5.7$ are $n=8.2\times10^{13}\,\rm cm^{-3}$, $r = 1.4\times10^9$\,cm. This further illustrates the issue with an AMD that is steep and rising to high $\xi$ values. Furthermore, this can indicate that some break in the AMD at high $\xi$ values is a fundamental property of such outflows in x-ray binaries. This issue is explored further in the discussion.


\begin{deluxetable}{lcccccccc}
\tabletypesize{\scriptsize}
\tablewidth{0pt}
\tablecaption{$n(r)$ parameters calculated for each observation based on Eq.\ref{eq_n0}}
\label{tab:density}
\tablehead{
\colhead{Observations ID} 
& \colhead{$\xi$ at $n_0$} 
& \colhead{$n_0$} 
& \colhead{$r_0$\,$^a$}
& \colhead{$r_{\rm peak}$\,$^b$}
& \colhead{$r_{\rm max}$\,$^c$}
& \colhead{$\alpha_1$}
& \colhead{$\alpha_2$}
& \colhead{$\alpha_3$}\\
\colhead{}
& \colhead{}
& \colhead{[\scriptsize{$\times10^{12}$\,cm$^{-3}$}]}
& \colhead{[\scriptsize{$\times10^{10}$\,cm}]}
& \colhead{[\scriptsize{$\times10^{10}$\,cm}]}
& \colhead{[\scriptsize{$\times10^{15}$\,cm}]}
& \colhead{}
& \colhead{}
& \colhead{}
}
\startdata
7485          & 6.5 & 2500         & 0.01           &            & 100       & 1.5           &     & \\
              &     & [1900--2600] & [0.009--0.011] &            & [10--200] & [1.44--1.52]  &     & \\[1.5ex]
23435+24663   & 3.4 & 5.8          & 2.6            & 4.7        & 3.0       & -2            & 1.5 & \\
              &     & [5.6--8.2]   & [1.7--3.5]     & [4--4.8]   & [0.9--40] & [-8 -- -0.67] & [1.44--1.6] & \\[1.5ex]
22213         & 3.3 & 5.5          & 1.7            & 3.6        & 9         & -4            & 1   & 1.667 \\
              &     & [5.3--7.0]   & [1.5--2.8]     & [3.2--3.7] & [1--30]   & [-10 -- -2]   &     & [1.6--1.7] \\
              &     &              &                & 7.2        &           &               &     & \\
              &     &              &                & [6.4--7.3] &           &               &     & \\
\enddata
\tablenotetext{a}{7485 - $\xi=10^{6.5}$, 23435+24663 - $\xi=10^{4.4}$, 22213 - $\xi=10^{4}$}
\tablenotetext{b}{23435+24663 - $\xi=10^{3.4}$, 22213 presents a peak plateau between two points - $\xi=10^{3.3},10^{3.0}$ }
\tablenotetext{c}{7485 - $\xi=10^{2}$, 23435+24663 - $\xi=10$, 22213 - $\xi=10$ }
\end{deluxetable}

\section{Discussion}

Reviewing the change in the AMD of an outflow over different outbursts of the same X-ray binary provides insights into the density profiles and location of the outflows. Although many works were published on \grs\, none delved deep into the ionization distribution of the wind or its density profile. 
Contrary to what we found for four previous XRB outflows \citep{Keshet2025}, where the shape of the AMD was steeply rising, some of the observations in this work show a dramatic turnover at high $\xi$. This is the first time we find a negative slope in an AMD which we interpret as an increasing $n(r)$, but only over a short range in $r$.  

Looking into the physical mechanism that could predict such a density inversion, where $n(r)$ is increasing, we think of a small-scale slab that is being compressed, possibly by radiation pressure. \citet{Stern2014} examined the scenario of radiation pressure compression in a small-scale slab of gas.
Their assumption of hydrostatic equilibrium implies that as photo-ionizing flux (pressure) penetrates the slab and is gradually absorbed, the gas pressure needs to increase. Since the temperature decreases as a result of loss of flux, the density must increase sharply as we go further into the slab.
This density profile naturally creates a broad, roughly flat, ionization distribution (AMD), unlike what we find here.
\citet{Stern2014} examined the dependency of their calculated AMD on total column density, metallicity, spectral slope of the ionizing source, and on $\xi _0$ at the slab surface. The AMD was relatively flat, independent of total column density and $\xi _0$. 
However, they tested only an ionizing source that is dominated by a power-law. For X-ray binaries in general, and especially as seen in this work by following the changes over different outbursts of \grs\, the main component of the SED is the disk black body. Therefore, we can not rule out a impact on the model of radiation pressure confinement for this scenario.  

In contrast, numerical simulations of thermally driven winds result in a complex, non-monotonic AMD. 
\citet{Dyda2017} found that thermal instabilities in winds of soft-state X-ray binaries (as those studied here) creates broad dips in the AMD below $\log \xi \approx 3$.
Between the dips the gas can be stable, giving the perception of an AMD turnover. At higher $\xi$'s the AMD drops as a power law, similar to our findings in \grs .
\citet{Waters2021} showed that thermal wind solutions in AGNs feature a drop in the AMD above $\log \xi \approx 3$ (see their Fig.\,13), see a similar feature in \citet[][Fig.\,5 therein]{Ganguly2021}. 
The AGN SED is again different from the X-ray binary case.
All of these thermal wind solutions produce a sharp increase in the AMD at $\log \xi < 2$, which is qualitatively different from the AMDs observed in X-ray binaries.

The density and launch radius we find for observation 7845 are within the ranges previously found by \citep{Miller2015, Zoghbi2016}. 
For observation 22213, our general description of the outflow is similar to what was found \citet{Miller2020, Neilsen2020} in terms of total column density, widths of the lines and outflow velocities. \citep{Miller2020} suggest two distinct absorption components, that perhaps indicate two separate outflows with different origins. It is possible that our assumption of a continuous outflow that results in a continuous AMD is wrong, and these are two separate outflows. One is at higher ionization and is a small-scale outflow, and the other a more "classic" wind spanning over a large scale and a wider range of ionization. In several outflows and different XRBs the H-like Fe was measured with different kinematics, that alludes to a different zone of the wind. The radii we find agree with the upper limits calculated for the two pion zones.

The density we measure agree with the predictions made by \citet{Neilsen2020}, of a density range $10^{12}-10^{13}\,\rm cm^{-3}$. The radii are smaller, since they predicted a few times $10^{11}\,$cm. This is only an approximation, since their model assumed a single power-law for the density as well as a single trend, unlike what we claim.  Interestingly, they did suggest a model with an increasing density profile, but this was assumed for the different areas of the outflow, meaning one trend to explain all parts of the wind. Our results indicate that a change in trend is necessary to explain the different ionic column density relations across all elements. Another prediction was that the density profile is most probably not a single power-law, which is consistent with what we find. Therefore it is clear all numbers quoted in this work cannot represent a full consistent solution, but rather presenting physical considerations to predict and justify order of magnitudes for the density and radius. Our calculation is more rigorous and attempts to give a full consistent solution for the wind. 

Our calculations from measurements of the density, radius and size of the wind are more robust than previously done in such outflows. They are directly calculated from the AMD, which is reconstructed based on line by line measurements and minimal model assumptions. As explained previously, the shape of the AMD is consistent even when exploring photoionization balance of different SEDs of the source.

\subsection{Discussion on the relation between density, radius and $\xi$}

If we examine the definition of the ionization parameter, $\xi=\frac{L}{nr^2}$, it is clear that a knowledge of $\xi$ and $L$ determines the quantity $nr^2$. Usually in X-ray diagnostics, disentangling $n$ and $r$ is difficult since there is no direct measurements of either one. Let us explore what can be extracted following some simplifying assumptions, for a general case of an outflow in an X-ray binary. We will begin with a more crude estimation and then elaborate further on a more detailed case. 
The typical luminosities for these binaries are $L\sim10^{37}-10^{38}$ erg. The ionization parameter relevant for the Fe K line region is usually $\log \xi=4-5$. The winds observed are measured with a typical \NHtot$\sim10^{23}-10^{24}\rm \, cm^{-2}$
First, let us say the total column density in the wind can be described by \NH$\simeq n\Delta r$. Further simplifying, we will take $\Delta r \sim r$. Therefore, we have two equations that determine the scale of $nr^2$:

\begin{equation}
    nr^2 = 10^{32}\left(\frac{L}{10^{37}}\right)\left(\frac{\xi}{10^{5}}\right)^{-1} = 10^{32}\, \rm cm^{-1}
\end{equation}

\begin{equation}
    10^{23}\,{\rm cm ^{-2}} =N_{\rm H}^{\rm tot} \simeq n r
\end{equation}

Together, these equations determine the possible $r$ and $n$ of the origin of the wind, assuming its origin coincide with the formation of the H-like and He-like Fe ions. The solution for the numbers presented is $r=10^9, n=10^{14}$. Writing a general expression for $n$ demonstrates it clearly


\begin{equation}
    n \simeq 10^{14}\left(\frac{L}{10^{37}}\right)^{-1}\left(\frac{\xi}{10^{5}}\right) \rm cm^{-3}
\end{equation}

Handling the expression for \NHtot\ more accurately, we assume the wind has a density profile of $n\sim r^{-\alpha}$. Integrating over this density profile over the width of the wind results in the measurement of \NHtot. For most of the outflows observed in binaries, there are lines identified from ions formed over several orders of magnitude in $\xi$. Therefore, we conclude the wind must be of a large scale. A small scale wind would require a very steep density profile to keep consistency with the start and end radii, as well as their appropriate ionization parameter values. This contradicts the AMD observed in such outflows. 

We will examine a density profile of the shape $n\sim r^{-1.5}$, which corresponds to an AMD comprised of a single power law with index $a=1$ (AMD$\sim\xi^a$). Notice that a shallower slope in the AMD, for example $a=0.5$, results in $\alpha=1.33$, which is not drastically different. As explained above, \NHtot\ is the result of integrating over the density profile along the width of the wind. As already mentioned, the wind is of a large scale. To simplify the integral we will take the end radius as infinity. 

\begin{equation}
    N_{\rm H}^{\rm tot} = n_0 \int_{r_0} ^\infty n(r)dr = n_0 \int_{r_0} ^\infty (\frac{r}{r_0})^{-1.5}dr = n_0[\frac{r^{-0.5}}{-0.5r_0^{-1.5}}]|_{r_0}^\infty \approx n_0 (\frac{r_0^{-0.5}}{-0.5r_0^{-1.5}})\approx n_0 r_0
\end{equation}

The ionization parameter and luminosity equation remain the same, and so we reach the same solution. The scale of the radius and density at the origin of the wind must be $r_0=10^9, n=10^{14}$. 
This radius, $r_0\sim10^{9}$\,cm, presents a physical problem. The velocities usually observed in these outflows are slower than the escape velocity expected at such a small radius.
This problem is most severe for the highest ionization species that are the closest to the center. 
This problem of high $\xi$ Fe-K dominated winds and their low velocities is not new and generic to X-ray binary outflows.

The maximal radii $r_{\rm max}$ obtained in Table\,\ref{tab:density} exceeds $7\times10^{12}$\,cm, the semi-major axis of the \grs\, binary. These extreme radii correspond to the lowest $\xi$ region that has a small fraction of the total column. Therefore the lowest $\xi$ and $r_{\rm max}$ are not well constrained. Winds extending beyond the binary separation are possible as long as the launching inclination is $\gtrsim10^\circ$ above the equatorial plane. This would imply higher velocities than we measure through Doppler shifts and decrease the tension with the low $r_0$ issue discussed above. 



In the detailed description above, for all cases the wind was assumed to be of a large scale. Let us discuss the case of a small scale wind. Since the maximal and minimal $\xi$ values are considered to be measured through the presence of absorption lines in the spectrum, the wind must contain a wide range of ionization levels. This results in a change of several orders of magnitude in $\xi$. Demanding a small scale wind, meaning $r_0\approx r_{\rm max}$, this results in an extremely steep density profile within a small slab. Taking the above common $L$ and assuming there are three orders of magnitude in $\xi$ between the launch radius and the outmost radius of the slab, the density profile must increase in three orders of magnitude within the slab. Assuming the slab length follows a conservative $\Delta r\sim r/3$, this results in a density slope of $n\sim r^{10}$ which is extremely steep. In stark contrast, a hydrostatically compressed slab results in a flat AMD \citep{Stern2014}.
In any event, a small scale absorber does not solve the generic tension between the closeness of the outflow to the center and the low velocities.
Conversely, the Keplerian periods at these radii of $10^{9} - 10^{10}$\,cm are 0.3 - 10\,s, at which QPOs have been detected in \grs\ \citep{Zhang2020}. Moreover, these radii correspond to $\sim 300 - 3000$ gravitational radii, which is commensurate with the broad line region in AGN, possibly implying this is the distance from which the outer thin accretion disk expands vertically. 

Exploration of the different assumptions and scenarios lead to an important conclusion - the ionization parameter and luminosity play a key role in determining the values of the density and radii of the wind. Furthermore, an ionization parameter that is very high, such as the case for the brightest \grs\, observation, result in a starting radius that is physicaly challenging. As a testing case, the density and radius calculation process was repeated for \fu\, which has a similar AMD with a lower breaking point \citep{Keshet2025}. There is a solution with $r_0\sim10^{9}$ cm, which is more probable physically. This means this is not an inherent problem with a steep AMD or the method, rather a reflection on the scales of the different wind parameters. 

\section{Conclusions}

In this paper, we explored the changes in the \grs\, outflow over different states. We focused on calculating the density profile, launch radius and geometrical extent of the wind based on the reconstructed AMD for four observations. The specific target also served as a starting point to a wider debate regarding these properties in X-ray binary outflows, and we reached the following conclusions

\begin{itemize}
    \item Reconstructing the AMD across multiple epochs reveals a structural difference between a continuous, steep power-law wind during soft high-flux states to a complex, broken distribution featuring distinct high-ionization turnovers during hard lower-flux states.
    \item The discovery of a negative AMD slope at $\log\xi > 3.3$ marks the first observational evidence of a localized density inversion ($n \sim r^{+\alpha}$) in an LMXB wind, indicating a highly compressed small-scale region. 
    \item For the hard states, we present direct, self-consistent calculations of the outflow properties, placing the wind density at $10^{12} - 10^{13}\text{ cm}^{-3}$ at a distance of $10^{10} - 10^{11}\text{ cm}$ from the central black hole. 
    \item In the soft ionizing state, high values of up to $\log\xi = 6.5$ are required to produce the He-like and H-like species, with no AMD turnover. This dictates a small launching radius of  $10^8\text{ cm}$ and a high density of $10^{15}\text{ cm}^{-3}$.
    \item Analytical modeling demonstrates that most generally, continuous, unbroken high-ionization AMDs dictate such small launch radii, which are in tension with the observed low outflow velocities. 
    The presence of high-ionization plateaus or breaks could fundamentally resolve this physical issue.
\end{itemize}

\begin{acknowledgments}
NK acknowledges the support of a Ramon scholarship from the Israeli Ministry of Science and Technology. This research was supported by The Israel Science Foundation (grant No. 2617/25). This paper employs a
list of Chandra data sets, obtained by the Chandra X-ray
Observatory, contained in~\dataset[DOI: 10.25574/cdc.650]{https://doi.org/10.25574/cdc.650}.

\end{acknowledgments}

\software{
          }



\bibliography{sample701}{}
\bibliographystyle{aasjournalv7}



\end{document}